\documentclass[12pt,a4paper,english]{article}
\usepackage[T1]{fontenc}
\usepackage[latin2]{inputenc}
\usepackage[english]{babel}
\usepackage{amsmath}
\usepackage{amsfonts}
\usepackage{indentfirst}
\usepackage[dvips]{graphicx}
\usepackage{youngtab}

\newcommand{\bea}{\begin{eqnarray}}
\newcommand{\eea}{\end{eqnarray}}
\newcommand{\be}{\begin{equation}}
\newcommand{\ee}{\end{equation}}

\newcommand{\hf}{\frac{1}{2}}

\newcommand{\C}{{\mathbb C}}

\begin{document}

\sloppy


\begin{flushright}
\begin{tabular}{l}
BONN-TH-2008-09\\

\\ [.3in]
\end{tabular}
\end{flushright}

\begin{center}
\Large{ \bf Deformed boson-fermion correspondence, Q-bosons, \\ and topological strings on the conifold}
\end{center}

\begin{center}

\bigskip

Piotr Su\l kowski

\bigskip

\medskip

\emph{Physikalisches Institut der Universit{\"a}t Bonn and Bethe Center for Theoretical Physics,} \\
\emph{Nussallee 12, 53115 Bonn, Germany} \\  [2mm]
\emph{and} \\ [2mm]
\emph{So{\l}tan Institute for Nuclear Studies, ul. Ho\.za 69, 00-681 Warsaw, Poland} \\ 

\bigskip

\emph{Piotr.Sulkowski@fuw.edu.pl}

\bigskip


\smallskip
 \vskip .4in \centerline{\bf Abstract}
\smallskip

\end{center}

We consider two different physical systems for which the basis of the Hilbert space can be parametrized 
by Young diagrams: free complex fermions and the phase model of strongly correlated bosons. Both systems 
have natural, well-known deformations parametrized by a parameter $Q$: the former one is related to 
the deformed boson-fermion correspondence introduced by N. Jing, while the latter is the so-called 
$Q$-boson, arising also in the context of quantum groups. These deformations are equivalent 
and can be realized in the same way in the algebra of Hall-Littlewood symmetric functions. Without a 
deformation, these reduce to Schur functions, which can be used to construct a generating function of 
plane partitions, reproducing a topological string partition function on
$\mathbb{C}^3$. We show that a deformation of both systems leads then to a
deformed generating function, which reproduces topological string 
partition function of the conifold, with the deformation parameter $Q$ identified with the size of 
$\mathbb{P}^1$. Similarly, a deformation of the fermion one-point function
results in the A-brane partition function on the conifold.


\newpage


\section{Introduction}

In this paper we consider two different physical systems with the same underlying structure: free 
complex fermions in two dimensions, and a chain of strongly interacting bosons. In both cases the 
relevant Hilbert spaces have a basis parametrized by Young 
diagrams, and elements of these basis can be represented by Schur functions. In particular, in the case 
of free complex fermions the mapping to Schur functions is a part of the well-known boson-fermion 
correspondence \cite{jimbo-miwa,kac}. There is a similar relation in the chain of interacting bosons, 
which originates in a non-standard algebra they obey.

Here we will be mostly interested in deformations of the above systems. In the context of free fermions,
a particularly interesting class of such deformations is related to the classical boson-fermion 
correspondence. The deformation we are mainly concerned with was introduced by N. Jing 
\cite{Jing1,Jing2}. It maps the states in the fermionic Hilbert space to the Hall-Littlewood symmetric 
polynomials $Q_{\lambda}$, which are a one-parameter generalization of Schur functions. One can also 
introduce the vertex operators $\Gamma_{\pm}(y)$, which acting on the vacuum $|0\rangle$ generate states 
$|\lambda\rangle$ corresponding to Young diagrams $\lambda$, and the coefficients of this expansion turn 
of to be the second species of Hall-Littlewood functions $P_{\lambda}$
\be
\prod_{j=1}^N \Gamma_-(y_j) |0\rangle = \sum_{\lambda} P_{\lambda}(y_1,\ldots,y_N;Q) |\lambda\rangle.  
\label{bosfer-intro}
\ee
In general there are more generalizations of the boson-fermion correspondence which are related to other 
families of symmetric functions \cite{Lam}.

The other system we analyze is the so-called $Q$-boson model, describing strongly interacting bosons on 
a chain \cite{BKT,BIK}. The $Q$-boson model is an integrable system, which can be solved within the 
framework of the Quantum Inverse Scattering Method \cite{BIK,QISM}. The algebra underlying the $Q$-boson 
model is more complicated than the standard bosonic algebra, and it arises also in the context of 
quantum groups \cite{kul-dem}. This system has an interesting limit of infinitely strong coupling, which 
corresponds to $Q=0$. This limit is called a \emph{phase model}, which is also the so-called crystal 
limit of the quantum groups \cite{kashivara}. The basis of the Fock space of the $Q$-boson model, and in 
particular its phase model limit, can also be parametrized by Young diagrams. Due to particular 
properties of the phase model algebra, this Fock space can also be represented by Schur functions, 
similarly as is the case for free complex fermions \cite{bogoliubov}. It then turns out that $Q$-boson 
model can be realized in the space of symmetric functions also in such way, that its states are mapped 
to the Hall-Littlewood polynomials \cite{tsilevich}. We discuss how, in a particular realization 
(which differs from the one in \cite{tsilevich} by normalization of states), 
the Hall-Littlewood polynomials in question are 
precisely $Q_{\lambda}$ which also arise in the context of the deformed boson-fermion correspondence. 
This relation allows to identify, in the limit of infinite $Q$-boson chain, the states of the deformed 
free fermions with those of the $Q$-boson model. In particular, in the framework of the Quantum Inverse 
Scattering Method one introduces certain creation operators $B(u)$ which acting on the vacuum generate 
$Q$-boson states corresponding to partitions $|\mu\rangle$, with coefficients also given by the 
Hall-Littlewood polynomials
\be
\prod_{j=1}^N B(u_j) |0\rangle = \sum_{\mu} P_{\mu}(u_1^2,\ldots,n_N^2;Q) |\mu\rangle. 
\label{Psi-P-intro}
\ee
In the limit of infinite chain the right sides of (\ref{bosfer-intro}) and (\ref{Psi-P-intro}) are the 
same, and (taking into account the subtlety concerning the zero-energy states) we can identify the two 
systems. In particular the vertex operators $\Gamma_-(y^2)$ are mapped to the creation operators $B(u)$.

Both systems mentioned above can also be used to compute generating functions of plane partitions of 
various shape. For free fermions (without any deformation) the counting is performed in terms of the 
vertex operators $\Gamma_{\pm}^{Q=0}(y_i)$ with a deformation parameter $Q=0$ and by specializing the 
values of $y_i$ to certain values \cite{ok-re,ok-re-Pearcey,ok-re-va}. These generating functions arise 
as overlaps of states of the form (\ref{bosfer-intro}) (with $Q=0$). Similarly, generating functions of 
plane partitions can be found in the phase model \cite{bogoliubov,boxed-shigechi} as overlaps of states 
of the form (\ref{Psi-P-intro}) with $Q=0$ and a particular choice of $u_j$. Due to the connection 
between the topological string theory and the counting of plane partitions \cite{foam}, the generating 
functions obtained in this way turn out to be equal to the partition functions of topological strings on 
certain backgrounds \cite{ok-re-va,sa-va,okuda,cube}. In particular, plane partitions in the 
unrestricted octant of $\mathbb{Z}^3$  lead to the partition function of $\mathbb{C}^3$ given by the 
MacMahon function $M(q)$ 

In the present paper we generalize the counting of plane partitions to the case of the deformed systems. 
Our main observation is the fact that replacing, in the computation of the MacMahon function, the vertex 
operators (or respectively creation operators $B(u)$) by their deformed counterparts, one obtains the 
partition function of the topological string on the resolved conifold. The deformation parameter $Q$ is 
then identified with $e^{-t}$, where $t$ is the K{\"a}hler parameter of the conifold. Similarly, the 
fermion one-point function generalizes from those of the A-brane in $\mathbb{C}^3$ \cite{sa-va} to the 
one of the A-brane in the resolved conifold. It is therefore quite interesting that some natural 
deformations of the three seemingly unrelated systems -- free fermions, strongly correlated boson, and 
topological strings -- are in a sense the same. 

The paper is organized as follows. In section \ref{sec-bos-fer} we review the deformed boson-fermion 
correspondence and its realization in the space of symmetric functions in terms of Hall-Littlewood 
polynomials. In section \ref{sec-phase} we introduce the phase model. In section \ref{sec-Qbos} we 
discuss its deformation to the $Q$-boson model, as well as its realization in the algebra of 
symmetric functions in terms of the same Hall-Littlewood polynomials as the deformed free fermions. In 
section \ref{sec-top} we discuss how free fermions or phase model can be used to compute generating 
functions of plane partitions, how they relate to the topological strings on $\mathbb{C}^3$, and how the 
deformation of both systems leads to the topological strings on the conifold. A short review of a theory 
of symmetric functions and in particular Hall-Littlewood polynomials is given in the appendix.


\section{Deformed boson-fermion correspondence} \label{sec-bos-fer}

Let us recall first the construction of the deformed boson-fermion correspondence \cite{Jing1,Jing2}. We 
consider the infinite-dimensional Heisenberg algebra generated by
\be
[\alpha_m, \alpha_n ] = \frac{m}{1-Q^{|m|}} \delta_{m,-n}.   \label{Q-commute-bos}
\ee
One then constructs generalized fermionic fields
\bea
\psi (z) & = & \Gamma_-(z) \Gamma_+(z)^{-1} e^{iz_0} z^{\alpha_0}, \nonumber \\ 
\psi^*(z) & = & \Gamma_-(z)^{-1}\Gamma_+(z) e^{-iz_0} z^{-\alpha_0},
\eea 
which are expressed in terms of the vertex operators
\be
\Gamma_{\pm}(z) = \exp \Big( \sum_{n\geq 1} \frac{1-Q^n}{n}\alpha_{\pm n} z^{\mp n} \Big).  
\label{gamma-pm}
\ee
From the Campbell-Hausdorff formula we find that these satisfy the commutation relation
\be
\Gamma_{+}(z) \Gamma_{-}(w) = \Gamma_{-}(w) \Gamma_{+}(z) \frac{w-Qz}{w-z}.   \label{commute-gamma}
\ee

We also define the modes $\psi_r, \psi_r^*$ by
\be
\psi(z) = \sum_{r\in\mathbb{Z}+\hf} \psi_r z^{-r-1/2},\qquad \psi^*(z) = \sum_{r\in\mathbb{Z}+\hf} 
\psi_r^* z^{-r-1/2}.
\ee
These modes satisfy the commutation relations
\bea
\{ \psi_r, \psi_s \} & = & Q \big(\psi_{r-1}\psi_{s+1} + \psi_{s-1}\psi_{r+1},   \big) \nonumber \\
\{ \psi^*_r, \psi^*_s \} & = & Q \big(\psi^*_{r-1}\psi^*_{s+1} + \psi^*_{s-1}\psi^*_{r+1},   \big) 
\label{Q-commute} \\
\{ \psi_r, \psi^*_s \} & = & Q \big(\psi_{r+1}\psi^*_{s-1} + \psi^*_{s+1}\psi_{r-1} \big) + 
(1-Q)^2\delta_{r,-s}.   \nonumber
\eea

There is the vacuum state $|0\rangle$ annihilated by all the positive modes
$$
\psi_r |0\rangle = \psi^*_r |0\rangle = 0,\qquad \textrm{for} \ r>0,
$$
as well as charged vacua
\bea
|m\rangle & = & \psi^*_{-m+1/2}\cdots \psi^{*}_{-3/2} \psi^{*}_{-1/2} |0\rangle, \qquad \textrm{for} \ 
m>0, \nonumber \\
|m\rangle & = & \psi_{-m+1/2}\cdots \psi_{-3/2} \psi_{-1/2} |0\rangle, \qquad \textrm{for} \ m<0. 
\nonumber
\eea

In the undeformed case there is a one-to-one correspondence between free fermion states and 
two-dimensional partitions \cite{kac}. In the neutral sector the state
\be
|\mu\rangle = \prod_{i=1}^{d} \psi^*_{-a_i-\hf} \psi_{-b_i-\hf}|0\rangle   \label{Rstate}
\ee
corresponds to the partition 
$$
\mu=(\mu_1,\ldots,\mu_l)
$$ 
with number of rows $l=l(\mu)$, such that
$$
a_i = \mu_i-i,\qquad b_i = \mu_i^t-i.
$$
The sequences $(a_i)$ and $(b_i)$ are necessarily strictly decreasing and they specify a partition in 
the so-called Frobenius notation 
\be
\mu = \left( \begin{array}{cccc} a_1 & a_2 & \ldots & a_{d(\mu)} \\
b_1 & b_2 & \ldots & b_{d(\mu)} \\
\end{array} \right),
\ee
where $d(\mu)$ denotes the number of boxes on a diagonal of a Young diagram of $\mu$. We often describe 
the partition also by specifying how many rows $n_i(\mu)$ of length $i$ it has, which is denoted by
\be
\mu = 1^{n_1(\mu)} 2^{n_2(\mu)} 3^{n_3(\mu)}\cdots    \label{R-i-ni}  
\ee

\begin{figure}[htb]
\begin{center}
\includegraphics[width=0.4\textwidth]{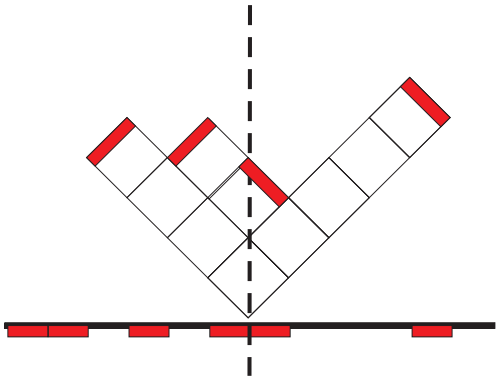}
\begin{quote}
\caption{A standard correspondence between partitions and states for $Q=0$. Positions of particles and 
holes are given by projecting the ends of the rows and the columns of a partition onto a Fermi sea. The 
partition drawn in the figure is $\mu=(5,2,2,1)\equiv 1^1 2^2 5^1$ and the corresponding state is 
$|\mu\rangle = \psi^*_{-\hf} \psi_{-\frac{3}{2}} \psi^*_{-\frac{9}{2}}\psi_{-\frac{7}{2}} |0\rangle$.} 
\label{fig-fermi-part}
\end{quote}
\end{center}
\end{figure}

It is easy to visualize this correspondence in terms of the Fermi sea. The vacuum $|0\rangle$ is given 
by a Fermi sea with all negative states filled and it is mapped to the trivial partition $\bullet$. A 
nontrivial partition is most easily visualized if one draws it with a corner fixed at the edge of the 
filled part of the Fermi sea. Then, the positions of particles and holes are read off by projecting the 
ends of the rows and the columns of this partition onto the Fermi sea, as shown in figure 
\ref{fig-fermi-part}. 

The above correspondence can be generalized to the deformed case with the help of Hall-Littlewood 
polynomials. Let us first identify the space of bosonic modes with the ring of symmetric functions by 
the mapping $\imath$ which associates $\alpha_{-n}$ to the Newton symmetric polynomial
$$
\imath(\alpha_{-n}) = p_n.
$$
This mapping is the isometric isomorphism. The main contribution of \cite{Jing1,Jing2} is the 
realization that this isometric isomorphism extends to the full deformed spaces of bosonic modes, and 
the images of fermionic states are Hall-Littlewood polynomials. In particular, the state obtained by 
application of $m$ creation operators $\psi^*_{-r_i}$ on the charged vacuum $|-m\rangle$ is mapped to
\be
\imath \big(|\mu\rangle\big) \equiv \imath \big(\psi^*_{-r_m}\cdots \psi^*_{-r_1}\big |-m\rangle) = 
Q_{\mu},    \label{HL-bos-fer}
\ee
where $Q_{\mu}$ is the Hall-Littlewood function (\ref{HL-Q}) associated to the partition 
$\mu=(\mu_1,\mu_2,\ldots)$, such that the sequence $(r_i)$ is decreasing and 
$$ 
\mu_i = r_i +i - \hf.
$$

The relations (\ref{HL-bos-fer}) and (\ref{sum-PQ}) imply that
\be
\prod_{j=1}^N \Gamma_-(y_j) |0\rangle = \sum_{\lambda} P_{\lambda}(y_1,\ldots,y_N;Q) |\lambda\rangle.   
\label{prod-gamma}
\ee

For $Q=0$ the relations (\ref{Q-commute-bos}-\ref{Q-commute}) and (\ref{prod-gamma}) reduce to those of 
the ordinary bosons and fermions, and the mapping (\ref{HL-bos-fer}) associates fermionic states to the 
Schur functions (\ref{schur}) in the standard way \cite{kac}
\be
Q=0\qquad \Rightarrow \qquad \imath \big(|\mu\rangle\big) \equiv \imath \big(\psi^*_{-r_m}\cdot 
\psi^*_{-r_1}\big |-m\rangle) = s_{\mu}.   
\label{bos-fer}
\ee


\section{Phase model}   \label{sec-phase}

In this section we consider a bosonic system based on the following algebra
\be
[N,\phi] = - \phi,\quad [N,\phi^{\dag}] = \phi^{\dag},\quad [\phi,\phi^{\dag}] = \pi,  \label{phase-alg}
\ee
with $\pi=|0\rangle \langle 0|$ the projection to the vacuum. The operator $\phi$ is one-sided isometry
$$
\phi \phi^{\dag} = 1,\qquad \phi^{\dag} \phi = 1-\pi.
$$
This algebra can be represented in the Fock space $\mathcal{F}$ consisting of $n$-particle states 
$|n\rangle$, such that
$$
\phi^{\dag}|n\rangle =|n+1\rangle,\quad \phi|n\rangle = |n-1\rangle, \quad \phi|0\rangle = 0, \quad 
N|n\rangle = n|n\rangle.
$$

The phase model is a model of a periodic chain with the hamiltonian \cite{BKT,BIK,bogoliubov}
\be
H = -\hf \sum_{i=0}^M \big(\phi^{\dag}_i \phi_{i+1} + \phi_i \phi^{\dag}_{i+1} -2N_i \big),   
\label{phase-ham}
\ee
with each set of operators $\phi_i, \phi^{\dag}_i, N_i$ satisfying the algebra (\ref{phase-alg}) and 
otherwise mutually commuting. The overall Fock space of the model is the tensor product of $M+1$ Fock 
spaces
\be
\mathcal{F} = \bigotimes_{i=0}^M \mathcal{F}_i.  \label{phase-fock}
\ee
The operator of the total number of particles is given by
$$
\hat{N} = \sum_{i=0}^M  N_i.
$$
The $N$-particle vectors in this space are of the form
\be
|\lambda\rangle = \bigotimes_{i=0}^M |n_i\rangle_i,\qquad \textrm{where} \ |n_i\rangle_i = 
(\phi_j^{\dag})^{n_j}|0\rangle_j,\quad  N=\sum_{i=0}^M n_i,    \label{Nparticle}
\ee
and we associate to it a partition $\lambda=1^{n_1} 2^{n_2} \ldots$. In fact, this association is not 
quite unique: the partition $\lambda$ itself does not know about the number $n_0$ of particles in the 
ground 
state $|0\rangle_i$. Nonetheless, if we fix the total number of particles $N$, we can deduce 
$n_0=N-n_1-n_2-\ldots = N-l(\lambda)$, where $l(\lambda)$ is the number rows in $\lambda$.

The phase model is integrable and it can be solved in the formalism of the Quantum Inverse Scattering 
Method \cite{QISM}. The solution is encoded in terms of the monodromy matrix
$$
T(u) = L_M(u) L_{M-1}(u)\cdots L_0(u),
$$
which is a product of L-matrices associated to each site of the chain
$$
L_i(u) = \left[ \begin{array}{cc}
u^{-1} & \phi^{\dag}_i \\
\phi_i & u 
\end{array} \right], \qquad i=0,\ldots, M,
$$
depending on the spectral parameter $u$. Each L-matrix, as well as the monodromy matrix, satisfies the 
intertwining relation
\bea
R(u,v)\big(L_i(u)\otimes L_i(v)  \big) & = & \big(L_i(v)\otimes L_i(u)  \big) R(u,v), \nonumber \\
R(u,v)\big(T(u)\otimes T(v)  \big) & = & \big(T(v)\otimes T(u)  \big) R(u,v), \label{intertwine}
\eea
with the $R$-matrix
\be
R(u,v) = \left[ \begin{array}{cccc}
f(v,u) & 0 & 0 & 0 \\
0 & g(v,u) & 1 & 0 \\
0 & 0 & g(v,u) & 0 \\
0 & 0 & 0 & f(v,u) 
\end{array} \right],   \label{Rmatrix} 
\ee
with 
$$ 
f(v,u) = \frac{u^2}{u^2-v^2},\quad g(v,u) = \frac{uv}{u^2-v^2}.
$$

The crucial objects in the following considerations are entries of the monodromy matrix, which we denote 
as
\be
T(u) = u^{-M} \left[ \begin{array}{cc}
A(u) & B(u) \\
C(u) & D(u) 
\end{array} \right],    \label{ABCD}
\ee
and which are operators acting in the Fock space (\ref{phase-fock}). In particular, the operators $B(u)$ 
and $C(u)$ are respectively creation and annihilation operators, in the sense that they increase and 
decrease the total number of particles
\be
\hat{N} B(u) = B(u) (\hat{N}+1), \qquad \hat{N} C(u) = C(u) (\hat{N}-1).   \label{creat-annih}
\ee
The operators $A(u)$ and $D(u)$ do not change the total number of particles. 

According to the Quantum Inverse Scattering Method, the eigenfunctions of the hamiltonian are of the 
form 
\be
|\Psi(u_1,\ldots,u_N) \rangle = \prod_{i=1}^N B(u_j) |0\rangle,   \label{Psi}
\ee
provided that the parameters $u_i$ satisfy the Bethe equations. Nonetheless, the states of this form are 
$N$-particle states, and may be of interest even if the Bethe equations are not satisfied. 

As shown in \cite{bogoliubov,tsilevich}, there is the following isometry between the states 
(\ref{Nparticle}) and the Schur functions  (\ref{schur})
\be
\jmath \Big(\bigotimes_{i=0}^M |n_i\rangle_i\Big) = s_{\lambda}, \qquad \textrm{with} \ \lambda = 
1^{n_1}2^{n_2}\ldots  \label{Qbos-schur}    
\ee
This relation implies that the states (\ref{Psi}) have the following expansion in the basis 
(\ref{Nparticle})
\be
|\Psi(u_1,\ldots,u_N) \rangle = \sum_{\lambda} s_{\lambda}(u_1^2,\ldots,u_N^2) \bigotimes_{i=0}^M 
|n_i\rangle_i,   \label{Psi-schur}
\ee
and the coefficients of this expansion are also the Schur functions. 

In the limit $M\to\infty$ the relation (\ref{Qbos-schur}) is the exact counterpart of the classical 
boson-fermion correspondence (\ref{bos-fer}), and we can identify the states of the phase model with 
those of the free fermion Fock space. Moreover, (\ref{Psi}) and (\ref{Psi-schur}) imply that the 
operator $B(u)$ can be identified with $\Gamma^{Q=0}_-(u^2)$, 
which is $Q=0$ limit of (\ref{gamma-pm}) \cite{bogoliubov,tsilevich}. Similarly, $C(u)$ can be 
identified with $\Gamma^{Q=0}_+(u^2)$. 
This is also the reason why the phase model can be used to compute the generating function of plane 
partition,
similarly as in \cite{ok-re-va}, by choosing the parameters $u_i$ appropriately. 
However, the phase model has an important advantage: when $M$ is finite, the operators $B(u)$ 
and $C(u)$ generalize $\Gamma^{Q=0}_{\pm}$, and are still manageable to manipulate, which allows to
compute explicitly the generating function of plane partitions in a box of finite height 
\cite{bogoliubov,boxed-shigechi}.


\section{Q-bosons}  \label{sec-Qbos}

There is a natural deformation of the phase model considered above. The algebra (\ref{phase-alg}) is the 
$Q=0$ limit of the so-called $Q$-boson algebra generated by operators $B, B^{\dag}$ and $N$  
\be
[N,B] = - B,\quad [N,B^{\dag}] = B^{\dag},\quad [B,B^{\dag}] = Q^N.  \label{Qbos-alg}
\ee
This algebra has been extensively studied e.g. in  \cite{BKT,BIK,QISM}, and it appears also in the 
context of quantum groups \cite{kashivara}.
We choose the following realization \footnote{it differs from the algebra in \cite{tsilevich} by the 
normalization of the state $|n\rangle$} of this algebra in the Fock space $\mathcal{F}$
\be
B^{\dag}|n\rangle = |n+1\rangle,\quad B|n\rangle = [n] |n-1\rangle, \quad B|0\rangle = 0, \quad 
N|n\rangle = n |n\rangle,    \label{Q-realize}
\ee 
with the scalar product given by
\be
\langle n | n \rangle = [n]!   \label{Q-product}
\ee
where we introduce the notation
\be
[n] = \frac{1-Q^n}{1-Q}, \qquad [n]! = \prod_{j=1}^n [j].  \label{Qbracket}
\ee

On the other hand, for $Q=1$ the $Q$-boson operators become ordinary bosons $B\to b$, $B^{\dag}\to 
b^{\dag}$, which satisfy $[b,b^{\dag}]=1$. 

Similarly as for the phase model, we consider the tensor product Fock space (\ref{phase-fock}) with 
$M+1$ components $\mathcal{F}_i$ and corresponding operators $B_j, B^{\dag}_j, N_j$. We again associate 
the
states in this Fock space with partitions (up to subtlety concerning the number of zero-energy particles 
$n_0$)
\be
|\lambda\rangle = \bigotimes_{i=0}^M |n_i\rangle_i.   \label{Qlambda}
\ee

From the scalar product (\ref{Q-product}) we get the norms of the $N$-particle states
\be
\langle \lambda | \lambda \rangle = \prod_{j=0}^M [n_j]! = 
\frac{\prod_{j=1}^{N-l(\lambda)}(1-Q^j)}{(1-Q)^N} b_{\lambda}(Q),
    \label{QQ-product}
\ee
with $b_{\lambda}(Q)$ defined as in (\ref{bQ}).

The generalization of the hamiltonian (\ref{phase-ham}) to the $Q$-boson case has the following form
\be
H = -\hf \sum_{i=0}^M \big(B^{\dag}_i B_{i+1} + B_i B^{\dag}_{i+1} -2N_i \big).   \label{Q-ham}
\ee
Writing 
\be
Q=e^{-\gamma}  \label{Q-gamma}
\ee 
with $0<\gamma \in \mathbb{R}$, the parameter $\gamma$ can be interpreted as a coupling constant 
associated to interacting terms arising in the expansion of $H$. Small coupling $\gamma$ corresponds to 
the free boson limit with the free hopping model hamiltonian. On the other hand, the limit of vanishing 
$Q$ corresponding to the phase model can be interpreted as the strong coupling limit with $\gamma\to 
\infty$ \cite{BIK}. 

Similarly as in the phase model, the solution of the $Q$-boson model is encoded in terms of the 
monodromy matrix
$$
T(u) = L_M(u) L_{M-1}(u)\cdots L_0(u) = u^{-M} \left[ \begin{array}{cc}
A(u) & B(u) \\
C(u) & D(u) 
\end{array} \right],
$$
with L-matrices of the form
$$
L_i(u) = \left[ \begin{array}{cc}
u^{-1} & B^{\dag}_i \\
(1-Q)B_i & u 
\end{array} \right], \qquad i=0,\ldots, M,
$$
depending on the spectral parameter $u$. L-matrices and the monodromy matrix satisfy the intertwining 
relation as in (\ref{intertwine}), but with the deformed $R$-matrix
$$
R_Q(u,v) = \left[ \begin{array}{cccc}
f(v,u) & 0 & 0 & 0 \\
0 & g(v,u) & Q^{-1/2} & 0 \\
0 & Q^{1/2} & g(v,u) & 0 \\
0 & 0 & 0 & f(v,u) 
\end{array} \right], 
$$
with 
$$ 
f(v,u) = \frac{Q^{-1/2}u^2-Q^{1/2}v^2}{u^2-v^2},\quad g(v,u) = \frac{uv (Q^{-1/2}-Q^{1/2})}{u^2-v^2}.
$$
This $R$-matrix is related to (\ref{Rmatrix}) by the limit $\lim_{Q\to 0} R_Q(u,v) = R(u,v)$.
The creation $B(u)$, annihilation $C(u)$, as well as $A(u)$ and $D(u)$ operators are defined as the 
components of the above monodromy matrix, in the same way as in (\ref{ABCD}). 

\subsection{Q-bosons and Hall-Littlewood polynomials}

We now extend the relation (\ref{Qbos-schur}) to the correspondence between the $Q$-boson state and 
Hall-Littlewood functions. In the realization (\ref{Q-realize}), the relevant functions are those given 
in (\ref{HL-Q})
\be
\jmath \Big(\bigotimes_{i=0}^M |n_i\rangle_i\Big) = Q_{\lambda}(x,Q), \qquad \textrm{with} \ \lambda = 
1^{n_1}2^{n_2}\ldots  \label{Qbos-HLQ}.    
\ee

Let us expand the creation operator as $B(u)=\sum_{k=0}^M u^{2k} b_k$. We show first (slightly modifying 
the proof in \cite{tsilevich}) that in the algebra of symmetric functions $b_k$ acts as the 
multiplication by $q_k$ given in (\ref{qk}). It is convenient to introduce the notation $B_j^{\dag} 
\equiv B_j^1, B_j \equiv B_j^{-1}, 1_j \equiv B_j^0$, so that 
$$
b_k  = \sum_{\epsilon_M,\ldots,\epsilon_0} (1-Q)^{\sum_i \delta_{-1,\epsilon_i}}  B_M^{\epsilon_M} 
\cdots B_0^{\epsilon_0},
$$
where $\epsilon_0\in\{0,1\}$, the highest non-vanishing $\epsilon_l=1$, $\epsilon_j\epsilon_{j+1}\neq 1$ 
and $\sum j\epsilon_j = k$. Acting on a state corresponding to the Hall-Littlewood polynomial $P_{\mu}$, 
the operator $B_j^1$ inserts one row of length $j$, while $B_j$ either removes one row of length $j$, or 
annihilates this state in case it did not contain any row of such length. This produces a state 
corresponding to certain partition $\lambda$. We therefore have $n_i(\lambda)=n_i(\mu)+\epsilon_i$, 
where the number of rows of length $i$ is given also by $n_i(\lambda) = \lambda^t_i-\lambda^t_{i+1}$. 
Introducing the skew diagram $\theta=\lambda-\mu$, we find 
$\theta^t_{i}=\theta_{i+1}^t+n_i(\lambda)-n_i(\mu) = \theta^t_{i+1} + \epsilon_i \in\{0,1 \}$. Because  
$\epsilon_j\epsilon_{j+1}\neq 1$, this means that $\theta$ is a horizontal strip (it has at most one box 
in each column), while the condition  $\sum j\epsilon_j = k$ implies that $\theta$ consists of $k$ 
boxes. Therefore $\theta$ is a horizontal $k$-strip and
$$
b_k P_{\mu} = \sum_{\lambda: \lambda/\mu \in \mathcal{H}_k} c(\mu,\lambda) P_{\lambda},
$$
and coefficients $c(\mu,\lambda)$ contain a factor $(1-Q)[n_j(\mu)]$ associated with each operator 
$B_j^{-1}$ and the realization (\ref{Q-realize}). Such factors arise for $\epsilon_j=-1$, which means 
that $\theta_j<\theta_{j+1}$, so that the set of such $j$'s is precisely the set $J$ introduced at the 
end of the Appendix. This way we get
$$
c(\mu,\lambda) = \prod_{j\in J} (1-Q^{n_j(\mu)}) = \psi_{\lambda/\mu}(Q)
$$
where the function $\psi_{\lambda/\mu}(Q)$ is given in (\ref{phipsi}). Finally, from Pieri formula 
(\ref{Pieri}) we see that $b_k$ indeed acts as a multiplication by the symmetric function $q_k$. 

Moreover, this means that the operator $B(u)$ corresponds to $\sum_{k=0}^M u^{2k} q_k(x_i)$, which can 
be treated as a specialization to finite number of variables of $\sum_{k=0}^M u^{2k} q_k(x_i)=Q(u^2)$ 
given in  (\ref{Q-gen-q}). Applying the operators $B(u_j)$ $N$ times and using (\ref{sum-PQ}) we get
$$
\jmath \Big(\prod_{j=1}^N B(u_j) |0\rangle \Big) = \prod_{i,j} \frac{1-Q x_i u_i^2}{1-x_iu_j^2}
= \sum_{\lambda} P_{\lambda}(u_1^2,\ldots,n_N^2;Q) Q_{\lambda}(x;Q),
$$
and therefore, with $|\lambda\rangle$ as given in (\ref{Qlambda})
\be
\prod_{j=1}^N B(u_j) |0\rangle = \sum_{\lambda} P_\lambda(u_1^2,\ldots,n_N^2;Q) |\lambda\rangle.        
\label{Psi-P}
\ee
This statement is the counterpart of the relation (\ref{prod-gamma}) in the deformed boson-fermion 
correspondence. The precise agreement we get in the limit $M\to \infty$; in particular, in this limit we 
can identify $Q$-boson operators $B(u)$ with the deformed vertex operators $\Gamma_-(u^2)$ 
(\ref{gamma-pm}). For finite $M$ the $Q$-boson model provides a generalization of the deformed 
boson-fermion correspondence.

\subsection{Examples} 

Let us fix $M=2$. By the straightforward expansion of the operators $B(u)$ into components we find
$$
B(u) = B_0^{\dag} + u^2(1-Q) B_0^{\dag} B_1 B_2^{\dag} + u^2 B_1^{\dag} + u^4 B_2^{\dag}.
$$
Applying three such operators to the vacuum we get the decomposition
$$
B(u_1) B(u_2) B(u_3) |0\rangle = \sum_{\lambda \subset [3,2]} P_{\lambda}|\lambda\rangle 
$$
where the sum runs over Young diagrams with at most 3 rows and 2 columns, and the coefficients are 
indeed Hall-Littlewood polynomials
\bea
P_{\bullet} (u_1^2,u_2^2,u_3^2;Q) & = & 1,\qquad \qquad \ \
P_{\ {\tiny \yng(1)}\ } (u_1^2,u_2^2,u_3^2;Q) =  u_1^2+u_2^2+u_3^2  \nonumber \\
P_{\ {\tiny \yng(2)}\ } (u_1^2,u_2^2,u_3^2;Q) & = & u_1^4+u_2^4+u_3^4 + 
(1-Q)(u_1^2u_2^2+u_1^2u_3^2+u_2^2u_3^2)  \nonumber \\
P_{\ {\tiny \yng(1,1)}\ } (u_1^2,u_2^2,u_3^2;Q) & = & u_1^2u_2^2+u_1^2u_3^2+u_2^2u_3^2  \nonumber \\
P_{\ {\tiny \yng(2,1)}\ } (u_1^2,u_2^2,u_3^2;Q) & = & u_1^2u_2^4+u_2^2u_1^4+u_1^2u_3^4 + 
u_3^2u_1^4+u_2^2u_3^4+u_3^2u_2^4 + (2-Q-Q^2)u_1^2u_2^2u_3^2  \nonumber \\
P_{\ {\tiny \yng(2,2)}\ } (u_1^2,u_2^2,u_3^2;Q) & = & u_1^4u_2^4+u_1^4u_3^4+u_2^4u_3^4 + 
(1-Q)u_1^2u_2^2u_3^2(u_1^2+u_2^2+u_3^2)  \nonumber \\
P_{\ {\tiny \yng(1,1,1)}\ } (u_1^2,u_2^2,u_3^2;Q) & = & u_1^2u_2^2u_3^2, \qquad
P_{\ {\tiny \yng(2,1,1)}\ } (u_1^2,u_2^2,u_3^2;Q) =  u_1^2u_2^2u_3^2(u_1^2+u_2^2+u_3^2)  \nonumber \\
P_{\ {\tiny \yng(2,2,2)}\ } (u_1^2,u_2^2,u_3^2;Q) & = & u_1^4u_2^4u_3^4, \qquad
P_{\ {\tiny \yng(2,2,1)}\ } (u_1^2,u_2^2,u_3^2;Q) = 
u_1^2u_2^2u_3^2(u_1^2u_2^2+u_1^2u_3^2+u_2^2u_3^2)  \nonumber 
\eea

For $M=3$ we get
$$
B(u) =  B_0^{\dag}+ u^2 B_1^{\dag} + u^4 B_2^{\dag}+u^6 B_3^{\dag} + u^2(1-Q) \big(B_0^{\dag} B_1 
B_2^{\dag} + B_0^{\dag} B_2 B_3^{\dag} \big) + u^4(1-Q) \big(B_0^{\dag} B_1 B_3^{\dag} + B_1^{\dag} B_2 
B_3^{\dag}  \big).
$$
The decomposition of the state
$$
B(u_1) B(u_2) |0\rangle = \sum_{\lambda \subset [2,3]} P_{\lambda}|\lambda\rangle 
$$
is also given by the sum runs over Young diagrams, this time with at most 2 rows and 3 columns, and the 
coefficients are the appropriate Hall-Littlewood polynomials
\bea
P_{\bullet} (u_1^2,u_2^2;Q) & = & 1,\qquad \quad \
P_{\ {\tiny \yng(1)}\ } (u_1^2,u_2^2;Q) =  u_1^2+u_2^2  \nonumber \\
P_{\ {\tiny \yng(2)}\ } (u_1^2,u_2^2;Q) & = & u_1^4+u_2^4 +(1-Q)u_1^2u_2^2  \nonumber \\
P_{\ {\tiny \yng(1,1)}\ } (u_1^2,u_2^2;Q) & = & u_1^2u_2^2, \qquad
P_{\ {\tiny \yng(2,1)}\ } (u_1^2,u_2^2;Q) =  u_1^2u_2^4+u_2^2u_1^4 \nonumber \\
P_{\ {\tiny \yng(2,2)}\ } (u_1^2,u_2^2;Q) & = & u_1^4u_2^4, \qquad 
P_{\ {\tiny \yng(3,2)}\ } (u_1^2,u_2^2;Q) = u_1^4u_2^6 + u_2^4u_1^6 \nonumber \\
P_{\ {\tiny \yng(3,1)}\ } (u_1^2,u_2^2;Q) & = &  u_1^2u_2^6 + u_2^2u_1^6 + (1-Q)u_1^4u_2^4  \nonumber \\
P_{\ {\tiny \yng(3,3)}\ } (u_1^2,u_2^2;Q) & = & u_1^6u_2^6, \qquad
P_{\ {\tiny \yng(3)}\ } (u_1^2,u_2^2;Q) = 
(1-Q)(u_1^2u_2^4+u_2^2u_1^4+u_1^6 + u_2^6)  \nonumber 
\eea


\section{Topological strings on the conifold}  \label{sec-top}

The A-model topological string partition function on $\C^3$ is given by the MacMahon function
$$
Z^{top}_{\C^3} = M(q)=\prod_{m,n\geq 1} \frac{1}{1-q^{n+m-1}}
$$
and it has been related to the topological vertex and the counting of plane partition in 
\cite{ok-re-va}. The generating function of plane partitions can be written as a fermionic correlator 
involving the standard vertex operators $\Gamma_{\pm}^{Q=0}(z_{n\pm})$ with a particular specialization 
of the values of 
$$
z_{n\pm}=q^{\pm(m-1/2)}.
 $$

Let us consider the same correlator as in \cite{ok-re-va} with the same specialization of $z$'s, but 
with vertex operators replaced by their deformed versions (\ref{gamma-pm})
\be
Z_Q = \langle 0 | \prod_{n\geq 1} \Gamma_+\big(q^{n-1/2}\big)\,\prod_{m\geq 1} 
\Gamma_-\big(q^{-(m-1/2)}\big)|0\rangle.   \label{coni-correlator}
\ee
We expect to get a modification of the $\C^3$ partition function.
Using the commutation relations (\ref{commute-gamma}), or the deformed 
boson-fermion correspondence (\ref{HL-bos-fer}) together with the 
scalar product in the algebra of symmetric functions 
(\ref{PQ-PP-QQ}), the above expression leads to
\be
Z_Q = \prod_{m,n\geq 1} \frac{1-Q q^{m+n-1}}{1-q^{m+n-1}} = \sum_{\mu,\nu} P_{\mu} P_{\nu} \langle \nu | 
\mu \rangle =  \sum_{\mu} P_{\mu} P_{\mu} b_{\mu}(Q) = Z^{top}_{conifold}(Q).  \label{Zconi-mod}
\ee
This reproduces the A-model topological string partition function on the resolved conifold, if we assume 
that 
$$ 
Q = e^{-t},
$$
where $t$ is the K{\"a}hler parameter of the conifold. Moreover, using the identification of the vertex 
operators $\Gamma_{\pm}$ with the $Q$-boson operators $B$ and $C$, the parameter $t$ can be identified 
with the $Q$-boson coupling constant (\ref{Q-gamma}). 


We could perform a similar calculation from the point of view of $Q$-bosons. From the statement 
(\ref{Psi-P}) we get a similar sum as in (\ref{Zconi-mod}), involving the scalar product $\langle \nu | 
\mu \rangle$, now with these states corresponding to $Q$-boson states. Nonetheless, there is a subtlety 
related to the number of zero-energy states $n_0$ mentioned earlier. In the $Q$-boson case the scalar 
product has the form (\ref{QQ-product}), so apart from the $b_{\mu}(Q)$ factor we are interested in 
there arise also some prefactors,
which can be discarded for infinite $N$.
However, for finite $N$ (which would correspond to counting partitions in a box of finite size), the 
correlator (\ref{coni-correlator}) with $\Gamma_{\pm}$ replaced by $Q$-boson operators $B$ and $C$ would 
lead explicitly to an answer which differs by these prefactors.

The relation between topological strings, plane partitions and fermions was also extended to include 
topological A-branes in \cite{sa-va}, where it was shown that a correlator of a fermion field 
$$
\psi^*_{\neq 0}(z) = \Gamma_-^{-1}(z)\Gamma_+(z)
$$
(with zero modes discarded) reproduces the open topological string partition function for the A-brane. 
We can repeat this computation in the case of the deformed operators, upon inserting $\psi^*_{\neq 0}$ 
(\ref{coni-correlator}). This yields
\bea
Z_Q & = & \langle 0 | \prod_{n\geq 1} \Gamma_+\big(q^{n-1/2}\big)\,\prod_{m=1}^{N+1}  
\Gamma_-\big(q^{-(m-1/2)}\big) \, \psi^*_{\neq 0}\big(q^{-(N+1/2)}\big) \prod_{m=N+2}^{\infty} 
\Gamma_-\big(q^{-(m-1/2)} \big) |0\rangle  = \nonumber \\
& = & Z^{top}_{conifold}(Q)\, \frac{L(a)}{L(aQ)}\, \xi(q,Q),
\eea
where $L(a)$ is the quantum dilogarithm (in the multiplicative notation) with the open string parameter 
$a=q^{N+1/2}$. This indeed reproduces the partition function of the A-brane, after discarding the factor 
$$
\xi(q,Q) = \prod_{m\geq 1} \frac{1-Qq^m}{1-q^m}.
$$
Similarly as before, this can be translated into the $Q$-boson language.


\section{Summary}  \label{sec-summary}

In this paper we reviewed two integrable models -- free fermions and strongly
correlated bosons -- and discussed their deformations, 
which can be realized in the same way in the algebra of symmetric functions in terms of 
Hall-Littlewood polynomials. Without a deformation, one can derive the generating functions of 
plane partitions using either of these systems, which reproduces the topological string partition 
function on $\mathbb{C}^3$. We showed that the deformation of both systems leads to a deformed partition functions, 
which coincide with the partition function of the resolved conifold, with or
without an A-brane. The K{\"a}hler parameter of the conifold is identified with a deformation parameter. 

First of all, there should be some deeper physical reasons why such different physical systems have 
deformations which are described by the same functions. In particular it would be interesting to realize 
the deformed boson-fermion correspondence and $Q$-bosons more explicitly in the context of topological 
strings. For example, the appearance of undeformed free fermions was related in \cite{dhsv} by a series 
of string theory dualities to a system of intersecting branes in the presence of the B-field. It would 
be nice to extend that picture to include the deformation considered in this paper. 

It is also tempting to understand whether underlying integrability of the strongly coupled bosonic chain 
could reveal some new features of the topological string theory, both in undeformed and deformed case.    

One could also generalize the present work in many directions. On one hand, there are more general 
deformation of the boson-fermion correspondence, related to other families of Schur functions 
\cite{Lam}. For example, replacing the commutation relations (\ref{Q-commute-bos}) by $[\alpha_m, 
\alpha_n ] = m\frac{1-T^{|m|}}{1-Q^{|m|}} \delta_{m,-n}$ leads to the Macdonald polynomials, which 
apparently appeared also in the context of topological strings and the so-called refined topological 
vertex \cite{gen-ver-jap, gen-ver}. On the other hand, on could generalize the computation of generating 
functions of plane partitions to more involved containers, and find their proper interpretation. 
Although such a computation is in principle possible in the free fermion framework \cite{okuda,cube}, 
the $Q$-boson model seems to be even better well-suited in this context 
\cite{bogoliubov,tsilevich,boxed-shigechi}.


\bigskip

\bigskip

\centerline{\Large{\bf Acknowledgments}}

\bigskip 

I would like to thank Robbert Dijkgraaf, Lotte Hollands, Ken Intriligator, Albrecht Klemm, Marcos Marino 
and Barry McCoy for useful discussions. I also thank the High Energy Physics group of the University of 
California San Diego for great hospitality during a part of this work. This project was supported by the 
Humboldt Fellowship.      

\vspace{0.5cm}



\appendix


\section*{\centerline{Appendix -- symmetric functions}} 

In this appendix we review a few properties of symmetric functions \cite{macdonald} which we need in our 
analysis. Let $\Lambda=\mathbb{Z}[x_1,x_2,\ldots]$ denote the ring of symmetric polynomials. This ring 
is graded 
with respect to the degree of a polynomial $k$
$$
\Lambda = \bigoplus_{k \geq 0} \Lambda_k.
$$
One can extend this ring by introducing an additional parameter $Q$, which leads to the ring 
$$
\Lambda \otimes \mathbb{Q}(Q)
$$
of symmetric functions over $\mathbb{Q}(t)$.

There are several useful basis of $\Lambda\otimes \mathbb{Q}(Q)$, which are parametrized by partitions. 
The Newton polynomial $p_{\mu} = \prod_l p_{\mu_l}$ are expressed by power sums $p_n=\sum_i x_i^n$. 
Monomial symmetric functions $m_{\mu}$ are sums of all distinct monomials obtained from $x^{\mu}=\prod 
x_i^{\mu_i}$ by permutations of $x_i$. Elementary symmetric functions $e_{\mu}=\prod e_{\mu_i}$ are 
determined in terms of the generating function $E(t) = \sum_{k\geq 0}e_k t^k = \prod_k (1+t x_k)$. 
Similarly, complete symmetric functions $h_{\mu}=\prod h_{\mu_i}$ are determined by the generating 
function $H(t) = \sum_{k\geq 0}h_k t^k = \prod_k (1-t x_k)^{-1}$. Schur functions are given as 
\be
s_{\mu} = \textrm{det} (h_{\mu_i-i+j}) = \textrm{det} (e_{\mu^t_i-i+j}).  \label{schur}
\ee 
One also introduces a scalar product on the space $\Lambda\otimes \mathbb{Q}(Q)$ by requiring that
\be
\langle p_{\mu}, p_{\nu} \rangle = z_R(Q) \delta_{\mu \nu},   \label{scalarprod}
\ee
where, using the notation (\ref{R-i-ni}),
\be
z_{\mu}(Q) = \frac{i^{n_i(\mu)} n_i(\mu)!}{\prod_{i\geq 1} (1-Q^{\mu_i})}.
\ee

We are particularly interested in Hall-Littlewood symmetric functions. There are two kinds of such 
functions, given by
\bea
P_{\mu}(x_1,\ldots,x_n; Q) & = & \frac{1}{v(Q)} \sum_{w\in S_n} w \Big(x_1^{\mu_1}\cdots 
x_n^{\mu_n}\prod_{i<j} \frac{x_i-tx_j}{x_i-x_j}  \Big),    \label{HL-P}  \\
Q_{\mu}(x_1,\ldots,x_n; Q) & = & b_{\mu}P_{\mu}(x_1,\ldots,x_n; Q) = \label{HL-Q} \\
& = & (1-Q)^{l(\mu)} \sum_{w\in S_n} w \Big(x_1^{\mu_1}\cdots x_n^{\mu_n}\prod_{i<j} 
\frac{x_i-tx_j}{x_i-x_j}  \Big),  \nonumber
\eea  
where
\bea
v_{\mu}(Q) & = & \prod_{i\geq 1} v_{n_i(\mu)}(Q), \qquad v_n(Q)=\prod_{j=1}^n \frac{1-Q^j}{1-Q}, 
\nonumber \\
b_{\mu}(Q) & = & \prod_{i\geq 1} \phi_{n_i(\mu)}(Q),\qquad \phi_n(Q) = \prod_{j=1}^n (1-Q^j).    
\label{bQ}
\eea
 
Hall-Littlewood symmetric functions interpolate between the Schur functions and the monomial symmetric 
functions,
$$
P_{\mu}(x;0) = s_{\mu}(x),\qquad P_{\mu}(x,1) = m_{\mu}(x). 
$$
The functions $P_{\mu}$ and $Q_{\mu}$ are dual to each other with respect to the scalar product 
(\ref{scalarprod}), so that
\be
\langle P_{\mu}, Q_{\nu} \rangle = \delta_{\mu \nu}, \qquad \langle P_{\mu}, P_{\nu} \rangle = 
\frac{\delta_{\mu \nu}}{b_{\mu}(Q)}, \qquad \langle Q_{\mu}, Q_{\nu} \rangle = b_{\mu}(Q)\,\delta_{\mu 
\nu},   \label{PQ-PP-QQ}
\ee
and they satisfy
\be
\sum_{\mu} P_{\mu}(x,Q)  Q_{\mu}(y,Q) = \sum_{\mu} b_{\mu}(Q) P_{\mu}(x,Q) P_{\mu}(y,Q) = \prod_{i,j} 
\frac{1-Qx_i y_j}{1-x_i y_j}.   \label{sum-PQ}
\ee

For a partition $\mu=(r)$ consisting of a single row of $r$ boxes, the generating function of
\be
q_r(x,Q) = Q_{(r)}(x,Q) = (1-Q) P_{(r)} (x,Q)   \label{qk}
\ee
is equal to
\be
Q(u) = \sum_{r=0}^{\infty} q_r(x,Q)u^r = \prod_i \frac{1-Qx_i u}{1-x_i u} = \frac{H(u)}{H(Qu)}.     
\label{Q-gen-q}
\ee

Finally, let $\mathcal{H}_r$ denote the horizontal $r$-strip, i.e. a skew partition $\theta=\mu / \nu$ 
whose columns consist of at most one box, i.e. $\theta'_i\in\{0,1\}$. Then we have the following Pieri 
formulas
\be
P_{\mu} q_r = \sum_{\lambda:\, \lambda/\nu\in\mathcal{H}_r} \varphi_{\lambda/\nu}(Q) P_{\lambda}, \qquad 
Q_{\mu} q_r = \sum_{\lambda:\, \lambda/\nu\in\mathcal{H}_r} \psi_{\lambda/\nu}(Q) Q_{\lambda},    
\label{Pieri}
\ee
where sums run over all diagrams $\lambda\supset \nu$ such that $\lambda/\nu$ is a horizontal $r$-strip, 
and
\be
\varphi_{\lambda/\nu}(Q) = \prod_{i\in I} (1-Q^{n_i(\lambda)}), \qquad \psi_{\lambda/\nu}(Q) = 
\prod_{j\in J} (1-Q^{n_i(\mu)}),     \label{phipsi}
\ee
where the set $I$ consists of integers $i\geq 1$ such that $\theta'_i>\theta'_{i+1}$ (so equivalently 
$\theta'_i=1$ and $\theta'_{i+1}=0$), while the set $J$ consists of integers $j\geq 1$ such that 
$\theta'_j<\theta'_{j+1}$ (so equivalently $\theta'_j=0$ and $\theta'_{j+1}=1$). One can verify that
$$
\varphi_{\lambda/\nu}(Q) / \psi_{\lambda/\nu}(Q) = b_{\lambda}(Q) / b_{\mu}(Q).
$$



\begin{thebibliography}{99}

\bibitem{jimbo-miwa}
M. Jimbo and T. Miwa,
\emph{Solitons and Infinite Dimensional Lie Algebras}, 
Kyoto University, RIMS 19 (1983) 943-1001.

\bibitem{kac}
V. G. Kac,
\emph{Infinite dimensional Lie algebras},
\textsf{Cambridge University Press 1990}.

\bibitem{Jing1}
N. Jing,
\emph{Vertex Operators and Hall-Littlewood Symmetric Functions},
Adv. in Math. 87 (1991) 226-248.

\bibitem{Jing2}
N. Jing,
\emph{Boson-fermion correspondence for Hall-Littlewood polynomials},
J. Math. Phys. 36 (1995) 12, 7073-7080.

\bibitem{Lam}
T. Lam,
\emph{A combinatorial generalization of the Boson-Fermion correspondence},
\textsf{math.CO/0507341}.

\bibitem{BKT}
N. M. Bogoliubov, R. Bullough, J. Timonen,
\emph{Critical behavior for correlated strongly coupled boson systems in 1+1 dimensions},
Phys. Rev. Lett. 25 (1994) 3933-3926.

\bibitem{BIK}
N. M. Bogoliubov, A. Izergin, N. Kitanine,
\emph{Correlation functions for a strongly correlated boson system},
Nucl. Phys. B 516 (1998) 501-528, \textsf{solv-int/9710002}. 

\bibitem{QISM}
V. Korepin, N. M. Bogoliubov, A. Izergin,
\emph{Quantum Inverse Scattering Method and Correlation Functions},
Cambridge University Press, 1993.

\bibitem{kul-dem}
P. Kulish, E. Damaskinsky,
\emph{On the q-oscillator and the quantum algebra $su_q(1,1)$},
J. Phys. A: Math. Gen. 23 (1990) L415.

\bibitem{kashivara}
M. Kashivara,
\emph{Crystalizing the q-analogue of universal enveloping algebras},
Commun. Math. Phys. 133 (1990) 249.



\bibitem{bogoliubov}
N. M. Bogoliubov,
\emph{Boxed Plane Partitions as an Exactly Solvable Boson Model},
\textsf{cond-mat/0503748}.

\bibitem{tsilevich}
N. Tsilevich,
\emph{Quantum inverse scattering method for the q-boson model and symmetric functions},
Funct. Anal. Appl. 40, No. 3 (2006) 207-217, \textsf{math-ph/0510073}. 

\bibitem{boxed-shigechi}
K. Shigechi, M. Uchiyama,
\emph{Boxed Skew Plane Partition and Integrable Phase Model},
J. Phys. A: Math. Gen. 38 (2005) 10287-10306, \textsf{cond-mat/0508090}.

\bibitem{ok-re}
A. Okounkov, N. Reshetikhin,
\emph{Correlation function of Schur process with application to local geometry of a random 3-dimensional 
Young diagram},
J. Amer. Math. Soc. 16 (2003) 581-603, \textsf{math.CO/0107056}.

\bibitem{ok-re-Pearcey}
A. Okounkov, N. Reshetikhin,
\emph{Random skew plane partitions and the Pearcey process},
 \textsf{math.CO/0503508}.

\bibitem{ok-re-va}
A. Okounkov, N. Reshetikhin, C. Vafa,
\emph{Quantum Calabi-Yau and Classical Crystals},
\textsf{hep-th/0309208}.

\bibitem{foam}
A.~Iqbal, N.~Nekrasov, A.~Okounkov and C.~Vafa,
\emph{Quantum foam and topological strings},
\textsf{hep-th/0312022}.

\bibitem{sa-va}
N. Saulina, C. Vafa,
\emph{D-branes as Defects in the Calabi-Yau Crystal},
\textsf{hep-th/0404246}.

\bibitem{okuda}
T. Okuda,
\emph{Derivation of Calabi-Yau Crystals from Chern-Simons Gauge theory},
JHEP 0503 (2005) 047, \textsf{hep-th/0409270}.

\bibitem{cube}
P. Su\l kowski,
\emph{Crystal Model for the Closed Topological Vertex Geometry},
JHEP 0612 (2006) 030, \textsf{hep-th/0606055}.

\bibitem{dhsv}
R.~Dijkgraaf, L.~Hollands, P. Su{\l}kowski and C.~Vafa,
\emph{Supersymmetric Gauge Theories, Intersecting Branes and Free Fermions},
JHEP 0802 (2008) 106, \textsf{hep-th/0709.4446}.

\bibitem{gen-ver-jap}
H. Awata, H. Kanno,
\emph{Instanton counting, Macdonald function and the moduli space of D-branes},
JHEP 0505 (2005) 039, \textsf{hep-th/0502061}.

\bibitem{gen-ver}
A. Iqbal, C. Kozcaz, C. Vafa,
\emph{The Refined Topological Vertex},
\textsf{hep-th/0701156}.

\bibitem{macdonald}
I. G. Macdonald,
\emph{Symmetric Functions and Hall Polynomials},
\textsf{Oxford Mathematical Monographs, 1995}.

\end{thebibliography}
\end{document}